\documentclass[aps]{revtex4}
\usepackage{epsf}
\usepackage{graphicx}

\begin{document}

\title
{Deterministic constant-temperature dynamics for dissipative quantum systems}

\author{Alessandro Sergi\footnote{E-mail: asergi@unime.it}}
\affiliation{
Dipartimento di Fisica, 
Universit\'a degli Studi di Messina,
Contrada Papardo 98166 Messina, Italy
}

\begin{abstract}
A novel method is introduced in order to treat the dissipative dynamics
of quantum systems interacting with a bath of classical degrees of freedom.
The method is based upon an extension of the Nos\`e-Hoover chain 
(constant temperature) dynamics to quantum-classical systems.
Both adiabatic and nonadiabatic numerical calculations on the relaxation dynamics
of the spin-boson model
show that the quantum-classical Nos\`e-Hoover chain dynamics
represents the thermal noise of the bath in an accurate and simple way.
Numerical comparisons, both with the constant energy calculation
and with the quantum-classical Brownian motion treatment of the bath,
show that the quantum-classical Nos\`e-Hoover Chain dynamics can 
be used to introduce dissipation in the evolution of a quantum subsystem
even with just one degree of freedom for the bath.
The algorithm
can be computationally advantageous in modeling, within computer simulation,
the dynamics of a quantum subsystem  interacting with
complex molecular environments.
\end{abstract}

\maketitle

One of the most natural ways to make a quantum system follow a dissipative
dynamics is achieved by putting it into contact with a thermal bath.
Since usually one is not interested in the detailed time evolution
of the bath degrees of freedom, it may also be convenient to approximate the
bath dynamics by means of a classical description.
When one faces with the problem of calculating the influence of
an environment  over a quantum subsystem,
this approach leads to the representation of (a certain class of)
open quantum systems~\cite{petruccione} by means of mixed quantum-classical theories.
Examples can be found in
many phenomena connected to quantum optics~\cite{qopt} and
quantum information theory~\cite{lebellac}. 
Typically, this is the case of cosmology where, 
due to the perduring lack of a full quantum theory
of gravitation, one is forced to approximate formalisms
in order to treat the interaction of quantum and classical degrees of freedom~\cite{wald}.
In many situations, condensed-matter quantum systems at finite temperature 
can also be treated  with mixed quantum-classical theories.
In light of the above discussion, one can certainly conclude that
mixed quantum-classical approximations
can be used in open quantum systems to describe 
many processes which are relevant to various fields of research.
It is worth noting that
mixed quantum-classical theories~\cite{qc-bracket} applied to condensed-matter
systems can treat classical molecular bath which can be
as complex as \emph{state-of-the-art} molecular dynamics
simulation techniques permit nowadays.

In the original constant-energy (NVE) formulation,
mixed quantum-classical algorithms require many environmental degrees of freedom in order
to describe the dissipative dynamics of the quantum subsystem.
This has been shown within a path-integral influence functional approach~\cite{makri}.
However, quantum-classical dynamics has been recently generalized~\cite{b3}
in order to be unified with the constant-temperature simulation
method originally developed by Nos\`e and Hoover~\cite{nose}
(more generally, the author in Ref.~\cite{b3} proposed a scheme in order to
unify quantum-classical dynamics with many energy-preserving phase space flows~\cite{bs}).
Therefore, one can think of using quantum-classical Nos\`e-Hoover (NH) dynamics
in order to describe dissipative effects and, in particular, 
the constant-temperature relaxation dynamics of a relevant quantum subsystem.
In practice, in order to overcome possible problems with ergodicity in classical
phase space, it is more convenient to generalize the Nos\`e-Hoover chain (NHC) method
of Martyna and coworkers~\cite{martyna} to the quantum-classical case
and to adopt it in place of the NH dynamics.
However, the choice of the NHC dynamics can be viewed
as a mere technical point with no deep
conceptual implication as far as quantum-classical theories are concerned.
In this Communication, 
by simulating the relaxation dynamics
of the spin-boson model~\cite{leggett}, I show that
the quantum-classical NHC dynamics can be adopted in order to describe 
dissipative effects in quantum-classical systems 
by means of a minimal (with respect to the number of degrees
of freedom explicitly taken into account) representation of the classical bath.

The quantum-classical Hamiltonian of the spin-boson model reads
\begin{eqnarray}
\hat{H}_{\rm sb}=-\hbar\Omega\hat{\sigma}_x+\sum_{j=1}^{N_b}
\left(\frac{P_j^2}{2M_J}+\frac{1}{2}M_j\omega_j^2R_j^2
-c_jR_j\hat{\sigma}_z\right)
\label{eq:Hsb}
\end{eqnarray}
where $2\hbar\Omega$ is the energy gap of the isolated two-state system,
$\hat{\sigma}_z$ and $\hat{\sigma}_x$ are Pauli's matrices,
$R_j$ and $P_j$ are the coordinates and momenta, respectively,
of $N_b$ harmonic oscillators with mass $M_j$ and frequencies $\omega_j$
making up the classical bath.
The other parameters of the system, \emph{i.e.}, $(M_j,\omega_j,c_j)$,
can be fixed by requiring that the harmonic bath is described by an Ohmic spectral density.
In order to study the relaxation dynamics of this model~\cite{qc-sb},
one can assume that the system is initially in an uncorrelated state
with the quantum subsystem in state $\vert\rm{up}\rangle$
and the classical harmonic bath in thermal equilibrium.
The corresponding quantum-classical density matrix
can be found starting from the full quantum expression by means
of a \emph{partial} Wigner transform~\cite{wigner}
and was explicitly written down in Ref.~\cite{qc-sb}.
The NVE quantum-classical dynamics of this system is formally exact~\cite{qc-sb}
(\emph{i.e.}, the quantum-classical equations of motion have the same form
that would arise within a full quantum treatment)
and numerical results, which agree very well with those obtained
by means of more sophisticated path-integral iterative techniques~\cite{makri},
are available in the literature~\cite{qc-sb}. 
Such NVE results, which were obtained with $N_b=200$, will be compared here
with those obtained with calculations performed
by means of the quantum-classical NHC dynamics and with the bath
made up by just one harmonic oscillator ($N_b=1$).
\begin{figure}
\resizebox{7cm}{4cm}{
\includegraphics* {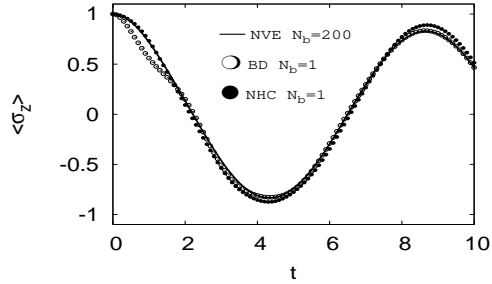}
}
\caption{Adiabatic dynamics of the spin-boson model:
$\beta=0.3$, $\Omega=1/3$, $\omega_{\rm max}=3$,
$\xi=0.007$. The continuous black line represents the NVE results
with $N_b=200$;
the white circles represent the BD results 
with $N_b=1$ and $\zeta=1$;
the black circles represent the NHC results with $N_b=1$.
}
\label{fig:fig1}
\end{figure}

For the spin-boson model, the quantum-classical NHC dynamics can be defined
upon introducing an extended Hamiltonian with
a chain of just two thermostat variables
\begin{equation}
\hat{H}_{\rm (NHC)}=\hat{H}_{\rm sb}+ \frac{p_{\eta_1}^2}{2m_{\eta_1}}
+\frac{p_{\eta_2}^2}{2m_{\eta_2}}+N_bk_BT\eta_1+k_BT\eta_2
\;,
\end{equation}
where $T$ is the temperature of the bath thermalizing the quantum spin,
$\eta_1$, $\eta_2$, $p_{\eta_1}$, $p_{\eta_2}$ are the Nos\`e variables, 
and $m_{\eta_1}$, $m_{\eta_2}$ are fictitious masses.
Following Ref.~\cite{b3}, the quantum-classical NHC bracket can be defined as:
\begin{eqnarray}
\left(\hat{H}_{\rm (NHC)}\right.&,&\left.\hat{\sigma}_z\right)_{\rm (NHC)}
=\frac{i}{\hbar}
\left[\begin{array}{cc}\hat{H}_{\rm (NHC)} & \hat{\sigma}_z\end{array}\right]
\nonumber \\
&\cdot&
\left[ \begin{array}{cc} 0 & 1+\frac{\hbar\Lambda^{\rm (NHC)}}{2i}\\
-1-\frac{\hbar\Lambda^{\rm (NHC)}}{2i} & 0\end{array}\right]
\cdot
\left[\begin{array}{c}\hat{H}_{\rm (NHC)}\\ \hat{\sigma}_z\end{array}\right]
\;,\nonumber\\
\label{eq:qcnosebracket}
\end{eqnarray}
where $\Lambda^{\rm (NHC)}$ is a bracket operator
whose action between two quantum-classical variables is defined as:
\begin{equation}
\hat{\xi}(X)\Lambda^{\rm (NHC)}\hat{\chi}(X)=-\sum_{IJ}
\frac{\partial\hat{\xi}}{\partial X_I} {\cal B}_{IJ}^{\rm (NHC)}
 \frac{\partial\hat{\chi}}{\partial X_J}\;.
\label{eq:noseop}
\end{equation}
Adopting as a convention for the point
of the extended Nos\`e phase space
$X=(R,\eta_1,\eta_2,P,p_{\eta_1},p_{\eta_2})$, the antisymmetric NHC matrix reads:
\begin{eqnarray}
\mbox{\boldmath$\cal B$}^{\rm (NHC)}
&=&\left[\begin{array}{cccccc} 0 & 0 & 0 & 1 & 0 & 0\\
0 & 0 & 0 & 0 & 1 & 0\\
0 & 0 & 0 & 0 & 0 & 1\\
-1 & 0 & 0  & 0 & -P & 0\\
0 & -1 &  0 & P & 0 & -p_{\eta_1}\\
0 & 0 & -1 & 0 & p_{\eta_1} & 0
\end{array}\right]\;.
\end{eqnarray}
It is worth noting that, since the Nos\`e coordinates are intrinsically classical,
a quantum-classical treatment of such a constant-temperature dynamics
is conceptually correct and, moreover, allows one to address nonadiabatic
effects. Other approaches~\cite{grillitosatti}, which do not use a quantum-classical bracket,
do not seem to permit nonadiabatic calculations in a straightforward manner.
The equations of motion can be written in the adiabatic basis as
\begin{equation}
\frac{d}{dt}\chi^{\alpha\alpha'}(X,t)
=\sum_{\beta\beta'}i{\cal L}_{\alpha\alpha',\beta\beta'}^{\rm (NHC)}
\chi^{\beta\beta'}(X,t)
\label{eq:qcnoseqofmadb}
\end{equation}
where
\begin{eqnarray}
i{\cal L}_{\alpha\alpha',\beta\beta'}^{\rm (NHC)}
&=&i\omega_{\alpha\alpha'}\delta_{\alpha\beta}\delta_{\alpha'\beta'}
+iL_{\alpha\alpha'}^{\rm (NHC)}\delta_{\alpha\beta}\delta_{\alpha'\beta'}
\nonumber\\
&-&J_{\alpha\alpha',\beta\beta'}
\\
iL_{\alpha\alpha'}^{\rm (NHC)}&=&\frac{1}{2}\sum_{IJ}{\cal B}_{IJ}^{\rm (NHC)}
\frac{\partial(H^{\alpha}_{\rm (NHC)}
+H^{\alpha'}_{\rm (NHC)})}{\partial X_J}\frac{\partial}{\partial X_I}
\;,\nonumber\\
\end{eqnarray}
and $\omega_{\alpha\alpha'}=(E_{\alpha}(R)-E_{\alpha'}(R))/\hbar$.
The quantum transitions operator, $J$,
is defined as in the constant-energy case~\cite{b3}.
\begin{figure}
\resizebox{7cm}{4cm}{
\includegraphics* {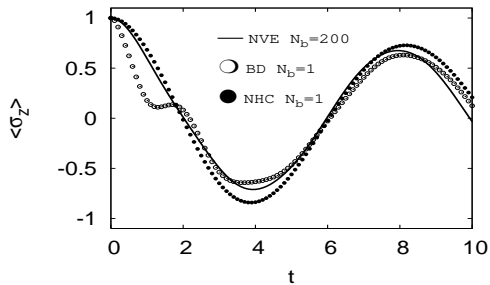}
}
\caption{Adiabatic dynamics of the spin-boson model:
$\beta=3$, $\Omega=1/3$, $\omega_{\rm max}=3$,
$\xi=0.1$. The continuous black line represents the 
NVE results with $N_b=200$;
the white circles represent the BD results
with $N_b=1$ and $\zeta=1$;
the black circles represent the NHC results with $N_b=1$.
}
\label{fig:fig2}
\end{figure}
One wants to calculate the time-dependent quantum-classical average: 
\begin{equation}
\langle\hat{\sigma}_z(X,t)\rangle
=\sum_{\alpha\alpha'}\int dX 
\rho^{\alpha'\alpha}_{W}
\sigma_z^{\alpha\alpha'}(X,t)
\label{eq:qcnoseave}
\end{equation}
where $\sigma^{\alpha\alpha'}_z(X,t)$ is given by Eq.~(\ref{eq:qcnoseqofmadb}).
Details of the numerical algorithm for calculating Eq.~(\ref{eq:qcnoseave})
both in the adiabatic and nonadiabatic limit, can be found in Ref.~\cite{qc-sb}. 
It is useful to recall that the nonadiabatic quantum-classical dynamics
can be pictured as a piece-wise deterministic propagation of the classical phase space point $X$
 over the energy surface 
$(\alpha\alpha')$ interspersed by stochastic quantum transitions
(realized by the action of $J$).
Note also that, in the NVE case, 
one must have $N_b\ge 200$,
as proven by a cumulant expansion analysis
of the influence functional entering the path-integral iterative procedure of Ref.~\cite{makri}.

In principle, dissipation  might also be described by means of
the quantum-classical Brownian dynamics (BD) that was introduced in Ref.~\cite{qcLangevin}.
In such a case,
the quantum-classical average of $\hat{\sigma}_z$ is still given by an equation
similar to~(\ref{eq:qcnoseave}) where, however, the time evolution is achieved
by means of a quantum-classical  Langevin-Liouville  operator,
whose explicit expression in the adiabatic basis
is known~\cite{qcLangevin}.
Such a stochastic operator is defined in terms of a friction constant, $\zeta$,
and of a Gaussian white noise process, $\xi(t)$, with the properties $\langle\xi(t)\rangle=0$, and 
$\langle\xi(t)\xi(t')\rangle=2k_BT\zeta\delta(t-t')$.
Therefore, it is interesting to check 
whether the Brownian dynamics of a bath with $N_b=1$
can also lead to an accurate dissipative dynamics. However, as shown in the following,
especially when considering
nonadiabatic effects, the numerical results prove that the NHC quantum-classical
dynamics provides a scheme which is much more accurate and robust than that arising from
the Brownian dynamics.

The spin-boson model has been simulated by using dimensionless coordinates~\cite{qc-sb}.
Hundred thousand trajectories were produced in order to sample the initial
condition in the nonadiabatic calculations~\cite{qc-sb}.
The system parameters 
were $\Omega=1/3$ and $\omega_{\rm max}=3$, while the Kondo parameter
and the reduced temperature took the two sets of values
($\xi=0.007$, $\beta=0.3$) and ($\xi=0.1$, $\beta=3$). The results 
of the NVE calculation,
with a bath composed of $N_b=200$ oscillators,
were compared with those obtained with just one oscillator ($N_b=1$)
in the two cases where either the NHC dynamics or the Brownian dynamics is used.
The outcome is that in the NHC dynamics a single oscillator
provides a good numerical representation of the dissipative quantum dynamics
of the spin. Moreover, it turns out that when nonadiabatic transitions are taken
into account, the quantum-classical NHC dynamics provides very good results
while the Brownian dynamics fails badly. Note that the the figures in this letter
display the results of BD calculations performed only with
$\zeta=1$ (in dimensionless units). However, various other calculations were performed,
with effective values of $\zeta$ between $0$ and $10$, without obtaining
any improvement in the nonadiabatic case.
\begin{figure}
\resizebox{7cm}{4cm}{
\includegraphics* {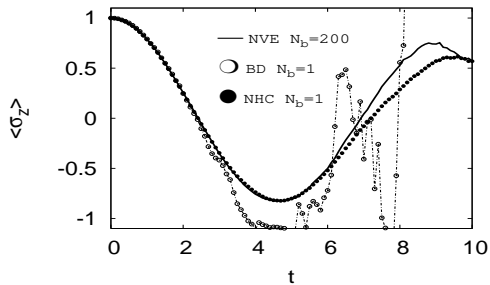}
}
\caption{Non-adiabatic dynamics of the spin-boson model including up to $6$ quantum transitions:
$\beta=0.3$, $\Omega=1/3$, $\omega_{\rm max}=3$,
$\xi=0.007$. 
The continuous black line represents the NVE result with $N_b=200$;
the black circles represent the NHC results with $N_b=1$;
the white circles (joined with a dashed line in order to help the eye)
represent the BD results with $N_b=1$ and $\zeta=1$.
}
\label{fig:fig3}
\end{figure}
Figures~\ref{fig:fig1}  and~\ref{fig:fig2} 
show the results of the adiabatic calculations 
for the set of parameters ($\xi=0.007$,$\beta=0.3$)
and  ($\xi=0.1$,$\beta=3$), respectively. In the adiabatic case, 
both the NHC and Brownian dynamics describe well the dissipative evolution
of the quantum subsystem interacting with a single oscillator.
The inclusion of nonadiabatic transitions (up to six for each trajectory in the 
quantum-classical ensemble)
shows that, with $N_b=1$, the NHC dynamics  is still very accurate while
the BD evolution becomes numerically unstable at short times.
This is not completely unexpected since when the system can switch from one
potential surface to another, because of the nonadiabatic transitions, the BD
dynamics (in the case of $N_b=1$) lacks of any equilibrating mechanism.
Instead, the quantum-classical NHC dynamics still
conserves the Hamiltonian along the trajectory. Such a conservation provides a robust
stabilization mechanism even for calculations 
with baths with very few degrees of freedom.
Figure~\ref{fig:fig3} shows the nonadiabatic results for the set of parameters
($\xi=0.007,\beta=0.3$) while those obtained with the set ($\xi=0.1,\beta=3$)
are displayed in Fig.~\ref{fig:fig4}.
In general, the results obtained
with the NHC nonadiabatic evolution
appear to be numerically more stable and smoother than those
obtained with NVE dynamics. This is even more
apparent in a slightly stronger coupling regime (see Fig.~\ref{fig:fig4}).

It must be remarked that surface-hopping calculations within 
nonadiabatic quantum-classical dynamics are, for the moment, limited
to relatively short times because of numerical instabilities~\cite{qc-sb}.
To address this issue, 
a quantum-classical non-linear formalism has been recently proposed~\cite{bwein}.
However, such long-time integration problems are not related to 
the NHC dynamics but challenge quantum-classical approximations 
of quantum dynamics on a more general level.
In order to clarify this point, Fig.~\ref{fig:fig5} displays the results
of a long-time calculation, performed in the adiabatic approximation.
Since there is a great interest in the phenomenon of driven
quantum tunneling~\cite{grifoni}, a static perturbation of the form 
$\hat{H}_{\rm ext}=-\hbar\gamma_s\hat{\sigma}_z$ was added to the
unperturbed spin-boson Hamiltonian in Eq.~(\ref{eq:Hsb}), and simulations
were carried out both in the NVE ($N_b=200$) and NHC case ($N_b=1$)
with $\gamma_s/\hbar=(1/3)\Omega$ while the other system parameters
took the same values as in the calculations whose results are illustrated
in Figs.~\ref{fig:fig1} and~\ref{fig:fig3}.
The results displayed in Fig.~\ref{fig:fig5} shows clearly that, in the adiabatic
approximation, the numerical agreement between the NVE dynamics ($N_b=200$)
and the NHC dynamics ($N_b=1$) is very good even over long time intervals.

The results provided in this Communication
suggest the possibility of representing the environmental noise,
leading to dissipative quantum dynamics, by means of 
deterministic NHC quantum-classical dynamics.
Such an idea can have deep conceptual implications since there is a 
subtle connection between thermal and quantum fluctuations~\cite{qft}.
Moreover, the algorithms presented here 
might open novel advantageous routes for the computer simulation of
quantum dynamics in open molecular environments.
\begin{figure}
\resizebox{7cm}{4cm}{
\includegraphics* {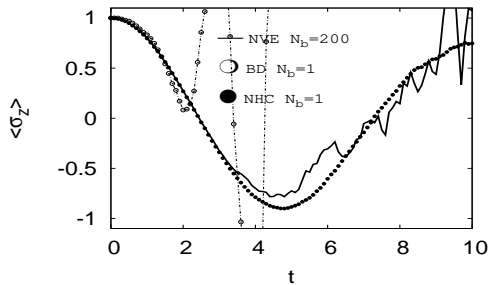}
}
\caption{Nonadiabatic dynamics of the spin-boson model
including up to $6$ quantum transitions:
$\beta=3$, $\Omega=1/3$, $\omega_{\rm max}=3$,
$\xi=0.1$. The continuous black line represents the
NVE results with $N_b=200$;
the black circles represent the NHC results with $N_b=1$;
the white circles (joined with a dashed line in order to help the eye)
represents the BD with $N_b=1$ and $\zeta=1$.
}
\label{fig:fig4}
\end{figure}
Within condensed matter systems, 
an example that might be studied by means of the approach illustrated
in this Communication is provided
by the system recently investigated in Ref.~\cite{miller}:
A retinal chromophore molecule evolving according to short-time quantum coherent dynamics 
in bacteriorhodopsin.
As already shown in Ref.~\cite{pyp}, a minimal model of such chromophore-protein
systems can be built by explicitly considering the chromophore molecule itself
and the nearest-neighbor
amino-acids, belonging to the tight-binding pocket in which the chromophore is contained.
On the short-time scale of the coherent quantum dynamics of the chromophore,
one might think of representing the dissipation entailed by the rest of the protein
by means of a deterministic NHC dynamics with a minimal bath. 
\begin{figure}
\resizebox{7cm}{4cm}{
\includegraphics* {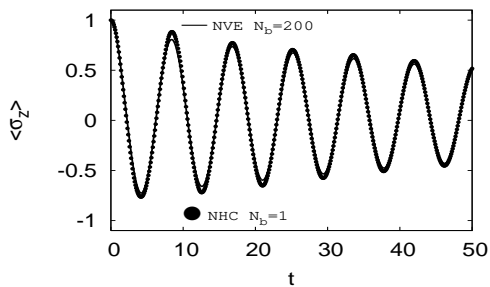}
}
\caption{Adiabatic dynamics of the spin-boson model
under the action of a static perturbation of the
form $\hat{H}_{\rm ext}=-\hbar\gamma_s\hat{\sigma}_z$:
$\beta=3$, $\Omega=1/3$, $\omega_{\rm max}=3$,
$\xi=0.1$, $\gamma_s/\hbar=(1/3)\Omega$. The continuous black line represents the
NVE results with $N_b=200$;
the black circles represent the NHC results with $N_b=1$.
}
\label{fig:fig5}
\end{figure}
It is also worth to note that there is a great interest in the field of quantum information
in the phenomenon of driven quantum tunneling~\cite{grifoni}.
In particular, recent work focuses on time-dependent external driving~\cite{grifoni2}.
In order to deal with such situations, one should generalize the
quantum-classical approach presented here in order to unify it with the techniques
of non-equilibrium molecular dynamics simulations.
Although non-trivial algorithmic issues might be expected,
this is something which is possible in principle and that deserves a
thorough future investigation.

In conclusion, the quantum-classical NHC dynamics may find interesting
applications in various fields: In fact, it may be used to simulate not only systems
in the fields of chemical-physics or biophysics but also in quantum-optics~\cite{qopt}
and quantum computing~\cite{lebellac}.

\vspace{0.5cm}
\noindent
{\bf Acknowledgments}\\
\noindent
I am grateful to Professor Paolo V. Giaquinta
and Dr. Giuseppe Pellicane for a critical reading of the manuscript.



\begin{thebibliography}{99}

\bibitem{petruccione}
H.-P. Breuer and F. Petruccione, The theory of open quantum systems
(Oxford University Press, Oxford, 2003).
\bibitem{qopt}
H. Paul, Introduction to Quantum Optics: From Light Quanta to Quantum Teleportation
(Cambridge University Press, Cambridge, 2004);
L. Mandel and E. Wolf, Optical Coherence and Quantum Optics  
(Cambridge University Press, Cambridge, 1995).
\bibitem{lebellac}
M. Le Bellac, A Short Introduction to Quantum Information and Quantum Computation
(Cambridge University Press, Cambridge, 2006).
\bibitem{wald}
R. M. Wald, Quantum Field Theory in Curved Spacetime and Black Hole Thermodynamics
(The University of Chicago, Chicago, 1994);
N. D. Birrell and P. C. W. Davies, Quantum fields in curved space
(Cambridge University Press, Cambridge, 1994).
\bibitem{qc-bracket}
I. V. Aleksandrov, Z. Naturforsch., {\bf 36a}, 902 (1981);
V. I. Gerasimenko, Theor. Math. Phys., {\bf 50},
77 (1982); D. Ya. Petrina, V. I. Gerasimenko and V. Z. Enolskii,
Sov. Phys. Dokl., {\bf 35}, 925 (1990);
W. Boucher and J. Traschen, Phys. Rev. D,
{\bf 37}, 3522 (1988);
W. Y. Zhang and R. Balescu, J. Plasma Phys., {\bf 40}, 199
(1988); R. Balescu and W.~Y. Zhang, J. Plasma Phys. {\bf 40}, 215
(1988); O. V. Prezhdo and V.V. Kisil, Phys. Rev. A, {\bf 56},
162 (1997); C. C. Martens and J.-Y. Fang, J. Chem. Phys.
{\bf 106}, 4918 (1996); A. Donoso and C. C. Martens, J. Phys.
Chem. {\bf 102}, 4291 (1998);
R. Kapral and G. Ciccotti,
J. Chem. Phys., {\bf 110}, 8919 (1999).
\bibitem{makri}
N. Makri and K. Thompson, Chem. Phys. Lett. {\bf 291}, 101 (1998);
K. Thompson and N. Makri, J. Chem. Phys. {\bf 110}, 1343 (1999);
N. Makri, J. Phys. Chem. B {\bf 103}, 2823 (1999).
\bibitem{b3}
A. Sergi, Phys. Rev. E {\bf 72} 066125 (2005).
\bibitem{nose}
S. Nos\`e, Mol. Phys. {\bf 52}, 255 (1984);
W. G. Hoover, Phys. Rev. A {\bf 31}, 1695 (1985);
S. Nos\`e,
Prog. Theor. Phys. {\bf 103}, 1 (1991).
\bibitem{bs}
A. Sergi, J. Chem. Phys. {\bf 124}, 024110 (2006);
Atti Accad. Pelorit. Pericol. Cl. Sci. Fis. Mat. Nat.
{\bf 33} c1a0501003 (2005);
Phys. Rev. E {\bf 72} 031104 (2005);
Phys. Rev. E {\bf 69} 021109 (2004);
Phys. Rev. E {\bf 67} 021101 (2003);
A. Sergi and M. Ferrario, Phys. Rev. E {\bf 64} 056125 (2001).
\bibitem{martyna}
G. J. Martyna, M. L. Klein, and M. Tuckerman, J. Chem. Phys. {\bf 92}, 2635 (1992).
\bibitem{leggett}
A. J. Leggett, S. Chakravarty, A. T. Dorsey, M. P. A. Fisher,
A. Garg, and W. Zwerger, Rev. Mod. Phys. {\bf 59}, 1 (1987).
\bibitem{qc-sb}
A. Sergi, D. Mac Kernan, G. Ciccotti, and R. Kapral,
Theor. Chem. Acc. {\bf 110} 49 (2003);
D. Mac Kernan, G. Ciccotti, and R. Kapral,
J. Chem. Phys. {\bf 116} 2346 (2002).
\bibitem{wigner}
E. P. Wigner, Phys. Rev. A {\bf 40} 749 (1932);
K. Imre, E. \"Ozimir, M. Rosenbaum, and P. Z. Zwiefel,
J. Math. Phys. {\bf 5} 1097 (1967);
M. Hillery, R. F. O'Connell, M. O. Scully, and E. P. Wigner,
Phys. Rep. {\bf 106} 121 (1984).
\bibitem{grillitosatti}
M. Grilli and E. Tosatti,
Phys. Rev. Lett. {\bf 62}, 2889 (1989).
\bibitem{qcLangevin}
A. Sergi and R. Kapral, J. Chem. Phys. {\bf 119}, 12776 (2003).
\bibitem{bwein}
A. Sergi,
J. Chem. Phys. {\bf 126}, 074109 (2007).
\bibitem{grifoni}
M. Grifoni and P. H\"anggi, Phys. Rep. {\bf 304}, 229 (1998).
\bibitem{qft}
J. Zinn-Justin,
Quantum Field Theory and Critical Phenomena
(Oxford University Press, Oxford, 1993);
M. Le Bellac, Quantum and Statistical Field Theory
(Clarendon Press, Oxford, 1991).
\bibitem{miller}
V. Prokhorenko, A. M. Nagy, S. A. Waschuk, L. S. Brown,
R. R. Birge, and R. J. Dwayne, Science {\bf 313}, 1257 (2006).
\bibitem{pyp}
A. Sergi, M. Gr\"uning, M. Ferrario and F. Buda,
J. Phys. Chem. {\bf 105}, 4386 (2001).
\bibitem{grifoni2}
M. C. Goorden, M. Thorwart, and M. Grifoni,
Phys. Rev. Lett. {\bf 93}, 267005 (2004);
Eur. Phys. J. B {\bf 45}, 405 (2005).



\end{thebibliography}
\end{document}